\documentclass[useAMS,usenatbib]{mn2e}
\usepackage[T1]{fontenc}

\usepackage{times}
\usepackage[english]{babel}
\usepackage{graphicx}
\usepackage{journals}
\usepackage{color}
\usepackage{amsmath}

\title{Metal enrichment signatures of the first stars on high-$z$ DLAs}

\author[Ma et al.]{Q. Ma$^{1,2,5}$, U.~Maio$^{3}$, B.~Ciardi$^1$, R.~Salvaterra$^4$\\
$^1$ Max-Planck-Institut f\"ur Astrophysik, Karl-Schwarzschild-Stra\ss e 1, D-85748 Garching bei M\"unchen, Germany\\
$^2$ Purple Mountain Observatory, Chinese Academy of Sciences, Nanjing 210008, China\\
$^3$ Leibniz-Institut f\"ur Astrophysik, An der Sternwarte 16, D-14482 Potsdam, Germany\\
$^4$ INAF, IASF Milano, via E. Bassini 15, I-20133 Milano, Italy\\
$^5$ University of Chinese Academy of Sciences, Beijing 100049, China
}

\begin{document}

\date{Accepted?? Received??; in original form??}


\pubyear{2016}

\maketitle

\label{firstpage}

\begin{abstract}
We use numerical N-body hydrodynamical simulations with varying PopIII stellar models to investigate the possibility of detecting first star signatures with observations of high-redshift damped Ly$\alpha$ absorbers (DLAs).
The simulations include atomic and molecular cooling, star formation, energy feedback and metal spreading due to the evolution of stars with a range of masses and metallicities.
Different initial mass functions (IMFs) and corresponding metal-dependent yields and lifetimes are adopted to model primordial stellar populations.
The DLAs in the simulations are selected according to either the local gas temperature (temperature selected) or the host mass (mass selected).
We find that 3\% (40\%) of mass (temperature) selected high-$z$ ($z\ge5.5$) DLAs retain signatures of pollution from PopIII stars, independently from the first star model. 
Such DLAs have low 
halo mass ($<10^{9.6}\,\rm M_{\odot}$),
metallicity ($<10^{-3}\,\rm Z_{\odot}$)
and star formation rate ($<10^{-1.5}\,\rm M_{\odot}\,yr^{-1}$).
{ Metal abundance ratios of DLAs imprinted in the spectra of QSO} can be useful tools to infer the properties of the polluting stellar generation and to constrain the first star mass ranges.
Comparing the abundance ratios derived from our simulations to those observed in DLAs at $z\ge5$, we find that most of these DLAs are consistent within errors with PopII stars dominated enrichment and strongly disfavor the pollution pattern of very massive first stars (i.e. 100~$\rm M_{\odot}$-500~$\rm M_{\odot}$). 
However, some of them could still result from the pollution of first stars in the mass range [0.1, 100]~$\rm M_{\odot}$.
In particular, we find that the abundance ratios from SDSS J1202+3235 are consistent with those expected from PopIII enrichment dominated by massive (but not extreme) first stars.
\end{abstract}

%
%
%
\section{Introduction}
\label{sec:intro}
One of the main challenges in studying primordial star formation is the lack of information on the first generation of structures.
The typical environment in which early star formation happens has been largely explored in the literature \cite[e.g.][and references therein]{Wise2012, Wise2014, Salvaterra2013, BM2013, Dayal2013, deSouza2013, deSouza2014,  Maio2016}, however the chemical details of the first galaxies remain elusive.

While direct observations of the first stars are very unlikely with current and planned instruments 
\footnote{
Very massive first stars, as predicted by several studies \cite[][]{Abel2002, Yoshida2006, Clark2011, Stacy2014, Ishiyama2016}, might be observable with the next generation space  telescope JWST \cite[for a review see][]{Bromm2013}, while primordial neutral gas could be detected by the future SKA mission \cite[][]{SKA2015}.
},
indirect signatures could be exploited to infer the properties of pristine stellar populations.
Precious tools in this respect are represented by the abundance patterns from (carbon-enhanced) extremely metal-poor stars \cite[][]{Spite2013} or gamma-ray bursts \cite[][]{Wang2012, MB2014, Ma2015, Ma2017}.
Indeed, stars with a range of masses and metallicities enrich the surrounding medium with different chemical signatures and on different timescales \cite[][]{Heger2002, Heger2010}.
Therefore, a deep investigation of atomic metal abundances in primordial gas clouds could unveil important properties of stellar populations at high redshift ($z$).

Techniques based on damped Ly$\alpha$ absorbers (DLAs) have also been proposed to detect the imprint of population~III (PopIII) stellar enrichment in their metal content \cite[][]{Salvadori2012, Maio2013, Kulkarni2013ApJ, Kulkarni2014ApJ}, in particular, in systems with halo masses smaller than $10^{9}\,\rm M_{\odot}$.
DLAs are dense gas absorbers with very high neutral hydrogen column density, i.e. $N_{\rm HI} > 2\times10^{20}{\rm cm}^{-2}$ \citep{Wolfe2005},
usually found along the line of sight of powerful QSOs.
Because of the large content of neutral material, they are expected to have a kinetic temperature below $10^4\,\rm K$ \cite[][]{Foltz1986, Smith1986, Wolfe1986} and to host significant amounts of molecules  \cite[][]{Noterdaeme2008}.
{ DLA systems are widely detected from $z=0$ to 7, while only one observed at $z\approx2.34$ \citep{Cooke2011} presents a carbon-enriched signature which might be associated to metal free supernovae of mass $\approx 25\,\rm M_{\odot}$ \citep{Kobayashi2011}.}
Recently, signatures of heavy elements have been detected in DLAs at $z \ge 5$
\cite[][]{Becker2012, Morrison2016}.
Their metal abundance ratios could be safely derived from ionic column densities.
Absolute metallicities cannot be measured directly, although they are expected to be less than 1\% solar \cite[][]{Maio2013, Keating2014, Keating2016, MT2015}.
Compared to local observations, e.g. metal-poor stars in the Milky Way, faint dwarf galaxies and low-$z$ DLAs, the detection of high-$z$ DLAs would provide stronger evidence of the first stars, thanks to the more significant PopIII contribution to the cosmic star formation rate density and metal enrichment at those epochs \cite[][]{Maio2010}.

In this paper, we use 3D numerical hydrodynamical chemistry simulations with different PopIII IMFs to study how gas pollution in primordial gas (in particular DLAs) changes with the properties of the first stars.
Despite former numerical studies based on
general lower-$z$ DLA properties
\cite[][]{Finlator2013, Finlator2015} or on semi-analytical modeling \cite[][]{Kulkarni2013ApJ, Kulkarni2014ApJ}, it has no attempt yet to investigate this issue by employing detailed 3D numerical chemistry simulations, as we plan to do here.

Throughout this paper, the standard flat $\Lambda$CDM cosmological model is adopted with the following parameters:
$\Omega_{0,\Lambda}=0.7$,
$\Omega_{0,M}=0.3$,
$\Omega_{b}=0.04$,
$h=0.7$,
$\sigma_8=0.9$
and $n=1$.
The simulations used are described in section~\ref{sec:simu}, the DLA definition and properties are presented in section~\ref{sec:dla}, the results for simulated metal abundances of galaxies are shown in section~\ref{sec:metal} and the conclusions are discussed in section~\ref{sec:discussion}.

\section{Simulations}
\label{sec:simu}

In this paper we employ the code extensively described in \cite{Maio2010, Maio2011, Maio2013t} and the simulations run in \cite{Ma2017}.
Here we mention only their main characteristics and we refer the reader to the original papers for more details and tests.
The simulations were performed by using an extended version of the GADGET-2 code \cite[][]{Springel2005} that includes atomic and molecular non-equilibrium chemistry, resonant and fine-structure cooling \cite[according to the implementation of][]{Maio2007}, star formation, stellar evolution and metal enrichment \cite[][]{Tornatore2007, Maio2010}.
All the simulations are started at $z=100$ and ended at $z=5.5$, and twenty snapshots from $z=17$ to 5.5 are saved.
We simulate $2 \times 320^3$ gas and dark matter particles in a box with side length $= 10 \, {\rm cMpc}\, h^{-1}$.
The initial mass is $3.39\times10^{5}\,\rm M_{\odot}$ for gas particles and $2.2\times10^{6}\,\rm M_{\odot}$ for dark matter particles.
We apply the same chemistry reactions and corresponding molecular and metal cooling of \cite{Maio2007}.
Stars are formed when, as a result of cooling, the density of a gas particle reaches $70\,\rm cm^{-3}$.
When this happens, a kinetic wind with velocity $500\,\rm km\,s^{−1}$ is added to the gas particle to prevent overcooling.
The stellar lifetimes we adopted are from \cite{Padovani_Matteucci1993}.
Stars ending their life as supernovae (SNe) \citep{Woosley1995,Heger2010} yield and spread abundant metals in the hosting galaxies.
As these metals mix quickly within the surrounding medium  \citep{Avillez2002}, in our simulations they are smoothed over the neighbours according to the SPH kernel \citep{Tornatore2007}.

Due to our current lack of definite information about stellar properties in primordial galaxies, in this work we employ simulations for three possible choices of the PopIII IMF: very massive SNe (VMSN), massive SNe (MSN) and regular SNe (RSN).
More specifically, in the simulation referred to as VMSN, the first stars are assumed to be in the mass range [100,~500]~${\rm M_\odot}$ with a Salpeter slope, and contribute to metal pollution { and thermal feedback} via pair-instability SNe (PISN) in the mass range [140,~260]~$\rm M_\odot$ \cite[][]{Heger2002}.
Both the MSN and RSN models sample a Salpeter IMF over [0.1,~100]~$\rm M_\odot$ for first stars, but they differ in the SNe ranges contributing to metal pollution, which are
[10,~100]~${\rm M}_{\odot}$ \cite[][]{Heger2010} and 
[10,~40]~${\rm  M}_{\odot}$ \cite[][]{Woosley1995, Heger2002}, respectively.
PopII/I stars are formed once the local gas has reached a metallicity $Z>10^{-4}~\rm Z_{\odot}$ (in solar units, $\rm Z_{\odot} \approx0.02$)
\cite[][]{BrommLoeb2003, Schneider2006}.
For PopII/I stars, we always adopt a Salpeter IMF in the range [0.1,~100]~$\rm M_\odot$
and include metal yields for AGB stars \cite[][]{vandenHoek1997},
type~Ia SNe (SNIa) \cite[][]{Thielemann_et_al_2003} and Type~II SNe (SNII) \cite[][]{Woosley1995}.
We assume that the energy released from all SNe is $10^{51}\,\rm erg$, with the exception of PISN, whose released energy is between $\sim 10^{51}$--$10^{53}\,\rm erg$ depending on the mass of their progenitors \cite[][]{Heger2002}.

Galaxies are identified by applying a friends-of-friends technique. 
In the following analysis, we only consider the galaxies with a minimum of 200 gas particles in order to prevent numerical artifacts \cite[][]{deSouza2013, Maio2013, Salvaterra2013}.
We also exclude metal-free galaxies to focus on the metal enrichment originated by PopIII and PopII/I pollution.

Following the procedure of \citet{Ma2017} to describe the level of metal enrichment from first stars, we define the fraction of metals in a galaxy which is produced by PopIII stars as:
\begin{equation}
\label{fiii}
f_{\mathrm{III}}
 =  \frac{ \sum_{j} m_{Z_j,{\rm III}} }{ \sum_{j} m_{Z_j} },
\end{equation}
where $m_{Z_j}$ is the mass of metal element $Z_j$ in the galaxy (i.e. except hydrogen and helium) and $m_{Z_j,{\rm III}}$ is the mass of metal element $Z_j$ coming from PopIII stars.
Then, we classify the selected galaxies in three classes based on the value of $f_{\mathrm{III}}$:
\begin{itemize}
\item[] 1: "PopII-dominated", if $f_{\rm III}<20\%$;
\item[] 2: "intermediate", if $20\%<f_{\rm III}<60\%$;
\item[] 3: "PopIII-dominated", if $f_{\rm III}>60\%$.
\end{itemize}

\begin{figure}
\centering
\includegraphics[width=0.90\linewidth]{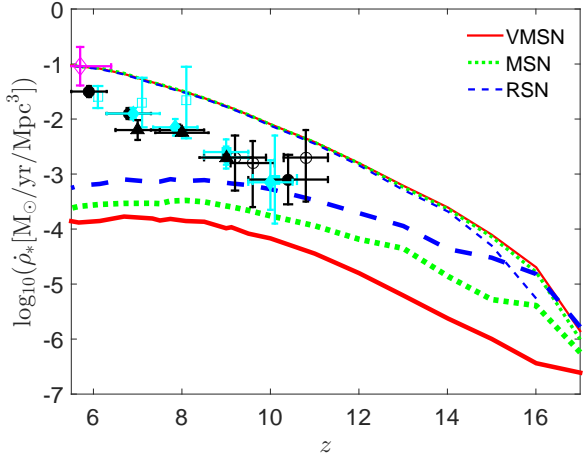}
\caption{
Redshift evolution of the intrinsic SFR comoving densities in model VMSN (solid red lines), MSN (dotted green) and RSN (dashed blue).
The lower set of thick lines denote the SFR densities of PopIII stars, while the upper set of thin lines are the SFR densities of PopII/I stars.
The data points with error bars show SFR densities derived by UV determinations from:
\citet{Bouwens2015} (filled black circles);
CLASH data in \citet{Zheng2012}, \citet{Coe2013} and  \citet{Bouwens2014}
(open black circles);
\citet{McLure2013,Ellis2013} (filled cyan diamonds);
\citet{Oesch2013,Oesch2014} (filled cyan squares);
\citet{Duncan2014} (empty cyan squares) and 
\citet{Laporte2016} (filled black triangles).
IR determination (empty magenta diamond) is from \citet{Rowan2016}.
\newline
(The color version is only available in the online journal.)}
\label{sfr_de}
\end{figure}
As a sanity check, we display in Fig.~\ref{sfr_de} the history of the comoving star formation rate (SFR) densities, $\dot{\rho}_{\ast}$, for both PopIII and PopII/I stars, as derived from the three simulations considered.
We also over-plot observational determinations inferred from data in different bands \cite[][]{Zheng2012, Coe2013, McLure2013, Ellis2013, Oesch2013, Oesch2014, Duncan2014, Bouwens2014, Bouwens2015, Rowan2016}.
While the simulations have very similar PopII/I SFR, the contribution from PopIII stars differs as a consequence of the different PopIII modeling \citep{Maio2010}.
The PopII/I SFR densities are comparable to the one estimated from observations in the infrared (IR) at $z \approx 6$ \cite[][]{Rowan2016}, but higher than most of the data points derived from the UV magnitudes of galaxies at $5 < z < 11$ \cite[][]{Bouwens2014, Bouwens2015, Duncan2014}.
These latter values may give an incomplete picture, though, since UV determinations could underestimate the contribution of embedded star formation \cite[][]{Rowan2016} { by} up to 1~dex \cite[][]{KennicuttEvans2012, MadauDickinson2014}.
Despite the lower dust content of early galaxies in comparison to $z<3$ objects expected by \cite{Capak2015}, the critical issues on high-$z$ IR luminosities \cite[][]{Bouwens2016, Laporte2016} and the origin of dust grains \cite[][]{Mancini2015, Ferrara2016, Bocchio2016} still persist. Hence, dust extinction might be important already at $z \sim 7$ and could hide the existence of dusty, UV-faint galaxies at early epochs \cite[][]{Salvaterra2013, Mancini2016, Mancuso2016}.
{ We also note that the SFR at $z>10$ might be slightly underestimated, as shown by simulations run at higher resolution \cite[][]{Maio2010,  Campisi2011}.}

%
%
\section{Damped Lyman-Alpha Absorbers}
\label{sec:dla}
%
In the following, we first introduce the definition of a DLA in the simulations and then discuss the properties of PopIII/PopII-dominated DLAs as well as their evolution with redshift.
\subsection{DLA definition}
\label{sec:dla:dla}
%
DLA (sub-DLA) in the observation is defined as a structure with column density of neutral hydrogen
$N_{\rm HI} > 2 \times 10^{20}\,{\rm cm}^{-2}$ ($10^{19} < N_{\rm HI} < 2 \times 10^{20}\,{\rm cm}^{-2}$)
along the line of sight of QSOs \citep{Peroux2003, Wolfe2005}.
To correctly evaluate the ionization fraction of neutral hydrogen within the cosmic medium in numerical simulations, the radiative transfer of ionizing photons should be accounted for.
Given the ability of our implementation to capture gas physics and chemistry down to very low temperatures, here we prefer to select DLAs in the simulations following the procedure of \citet[][]{Maio2013} and \citet{MT2015}, i.e. gaseous objects with mass-weighted gas temperature $T \le 10^{4}\, \rm K$ \cite[consistently with][]{Cooke2015}. { Here $T$ is calculated as $\sum_{k} m_{k}T_{k}/\sum_{k} m_{k}$, where $k$ denotes all the gas particles in one halo, and $m_{k}$ and $T_{k}$ are the mass and temperature of the $k$-th gas particle, respectively.}

Alternatively, we also consider the description of DLAs in terms of a mass-dependent neutral hydrogen cross-section \cite[as done in][]{Kulkarni2013ApJ, Kulkarni2014ApJ}, which is parameterized through a fit calibrated against $z=3$ observations \cite[see also][]{Pontzen2008, FontRibera2012}:
\begin{equation}
\label{eq:fit}
\Sigma(M_{\rm halo}) =
\Sigma_0 \left( \frac{M_{\rm halo}}{M_0} \right)^{2} \left( 1+\frac{M_{\rm halo}}{M_0} \right)^{\alpha-2},
\end{equation}
where
$\alpha=0.2$, 
$M_0=10^{9.5}\,\mathrm{M}_{\odot}$ 
and $\Sigma_0=40\,\mathrm{kpc}^{2}$
at $z=3$.
This fitting function matches very well the column density distribution function observed at low $z$ \cite[][]{Kulkarni2013ApJ}.
However, we note that a possible bias may appear for high-$z$ DLAs, due to the fact that the fit is calibrated to reproduce data at $z = 3$.

To aid the assessment of the impact of such definitions on our results, we discuss below some general galactic properties.
\begin{figure}
\centering
\includegraphics[width=0.90\linewidth]{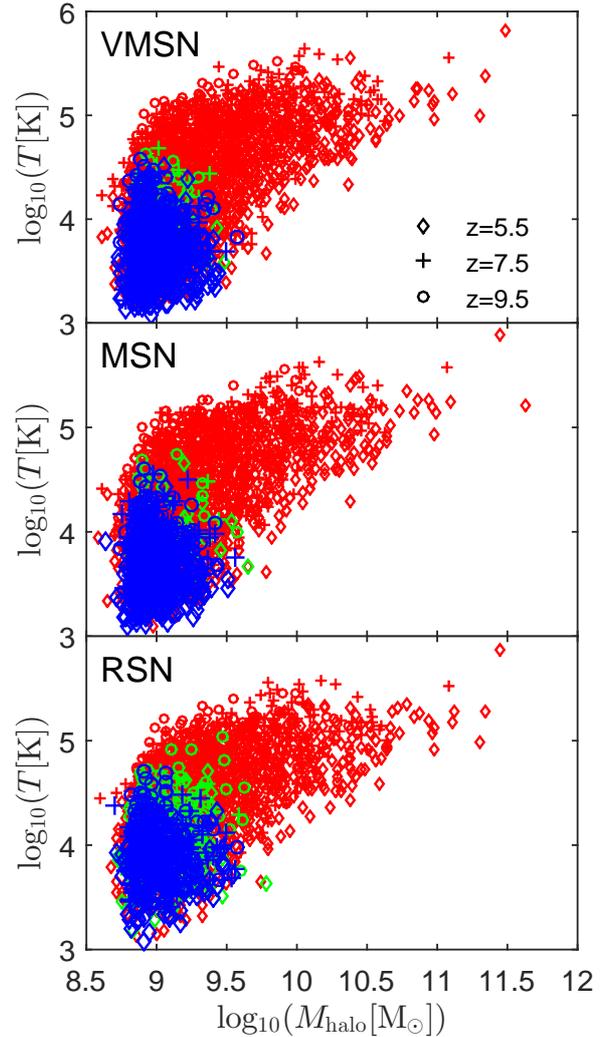}
\caption{
Distributions of mass-weighted temperature ($T$) of PopII-dominated (small red symbols), intermediate (green) and PopIII-dominated (big blue) galaxies as a function of halo mass ($M_{\rm halo}$). The results are shown at $z=9.5$ (circles), 7.5 (crosses) and 5.5 (diamonds). From top to bottom, the panels refer to the VMSN, MSN and RSN model.
\newline
(The color version is only available in the online journal.)
}
\label{tvsmgas}
\end{figure}
In Fig.~\ref{tvsmgas}, we present the distributions of PopII-dominated (red), intermediate (green) and PopIII-dominated (blue) galaxies at three redshifts in terms of mass-weighted temperature $T$ vs dark-matter halo mass $M_{\rm halo}$.
As expected, more massive galaxies appear with decreasing redshift. 
For example, at $z=5.5$ halo masses are between $10^{8.5}\, \rm M_{\odot}$ and $10^{11.5}\,\rm M_{\odot}$, with corresponding gas temperatures ranging from $10^3\,\rm K$ to several $10^5\,\rm K$, while at $z=9.5$ the halo masses are $< 10^{10}\,\rm M_{\odot}$.
The more massive galaxies usually have higher gas temperature, as the higher-mass objects are statistically more evolved and have deeper potential well, higher SFR and stellar mass.

No matter at which redshift, PopII-dominated galaxies show very similar $T$ vs $M_{\rm halo}$ distributions in the three simulations, i.e. their mass and temperature are weakly affected by the first star model adopted.

Also the distributions of PopIII-dominated galaxies are consistent in the three simulations, with halo masses
$M_{\rm halo}<10^{9.6}\, \rm M_{\odot}$ 
and temperatures $T<10^{4.5}\,\rm K$, 
lower than those of the PopII-dominated galaxies.
No significant redshift evolution is evident here, since PopIII metal pollution usually dominates in low-mass galaxies, where gas parcels feature mostly pristine composition, irrespectively from redshift.

The halo mass and gas temperature of intermediate galaxies are very close to those of PopIII-dominated galaxies, with a few rare exceptions.
The paucity of intermediate-class galaxies is due to the rapidity of metal enrichment, which is quickly dominated by PopII/I star formation (i.e. $f_{\rm III}<20\%$).
Due to the different stellar lifetimes, metal spreading from first PopIII SNe is delayed in the RSN model with respect to the VMSN and MSN models. Therefore, the PopIII SFR reaches higher levels and more prolonged duration \citep{Ma2017}, delaying the transition from intermediate to PopII-dominated galaxies (i.e. more galaxies retain $f_{\rm III}>20\%$).
For this reason, intermediate galaxies are more abundantly present in the RSN model than in the VMSN and MSN models.

From Fig.~\ref{tvsmgas} it is clear that the two theoretical definitions of DLAs are expected to affect the results on the predicted properties.
Indeed, since PopIII-dominated objects have very low gas temperatures, most of them could be DLA candidates when selected with the temperature threshold 
$T \le 10^4\,\rm K$ (in the following referred as "temperature selected"), while PopII-dominated ones, populating the high-mass end of the distributions, would dominate the candidate DLA population when selected according to eq.~\ref{eq:fit} ("mass selected").
The differences of these two approaches are also discussed in \cite{Morrison2016}.
In the following, the evolution and the statistical properties of DLAs selected with both criteria will be shown.

%
\subsection{Evolution of PopII/PopIII-dominated DLAs}
\label{sec:dla:evo}
%
\begin{figure}
\centering
\includegraphics[width=0.9\linewidth]{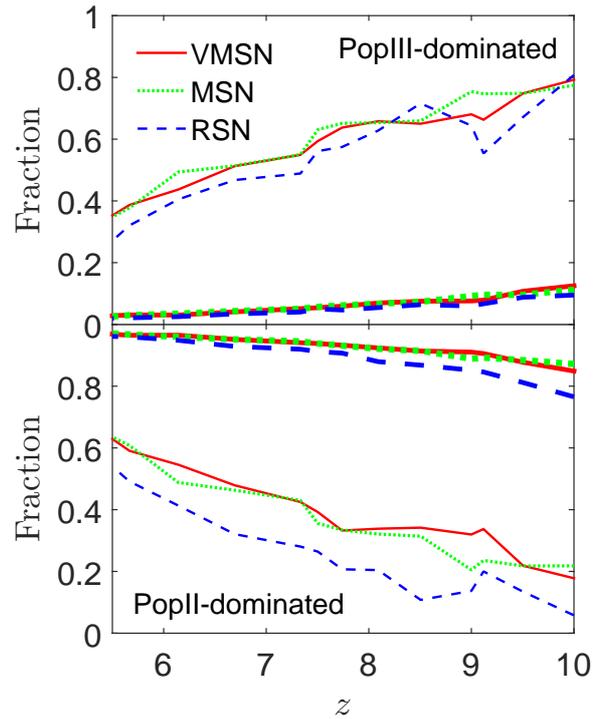}
\caption{
Redshift evolution of the fraction of DLAs which are PopIII-dominated (top) and PopII-dominated (bottom) in models VMSN (solid red lines), MSN (dotted green) and RSN (dashed blue).
The thin and thick sets of curves refer to temperature selected DLAs and mass selected DLAs, respectively.
\newline
(The color version is only available in the online journal.)
}
\label{lyafvz}
\end{figure}
In Fig.~\ref{lyafvz} we show the evolution of the fraction of PopIII-dominated (top panel) and PopII-dominated (bottom panel) DLAs as a function of redshift, according to both DLA definitions.
For temperature selected DLAs (thin lines), the fraction is defined as $N_{i}(z)/N_{\rm tot}(z)$, where $N_{i}(z)$ is the number of class $i$ DLAs and $N_{\rm tot}(z)$ is the total number of all classes DLAs at redshift $z$.
For mass selected DLAs (thick lines) the fraction of class $i$ is computed as $\Delta_{i}(z)/\Delta_{\rm tot}(z)$, where $\Delta_{i}(z)$ is sum of $\Sigma(M_{\rm halo})$ of class $i$ galaxies and $\Delta_{\rm tot}(z)$ is the integral of $\Sigma(M_{\rm halo})$ of all galaxies at redshift $z$. 

Consistently with what mentioned in the previous section, the fraction of PopIII-dominated DLAs decreases with redshift because of the increasing metal contribution from PopII/I stars.
{ As the cosmic metallicity is hardly affected by the mass range of the first stars \citep{Maio2010},} this fraction is almost independent from the first star models, although the RSN model shows slightly lower PopIII-dominated values than the others, as a result of the delayed metal pollution from smaller-mass first stars that have longer lifetimes.
On the other hand, while temperature selected DLAs have a very high probability to be PopIII-dominated, e.g. 80\% at $z=10$ and 40\% at $z=6$, the corresponding fraction for mass selected DLAs is only 10\% at $z=10$ and 3\% at $z=6$.
The difference is caused by the fact that PopIII-dominated galaxies are likely to have $T\le 10^{4}\,\rm K$ and small halo masses (see Fig.~\ref{tvsmgas}). These characteristics are strongly disfavoured when using the fit of eq.~\ref{eq:fit}
(which gives a larger weight to typically hotter objects with masses $\gtrsim 10^{9.5} \rm M_\odot$).

The fraction of PopII-dominated DLAs is complementary to that of PopIII-dominated ones.
It turns to be very high for mass selected DLAs, with fractions exceeding $80\%$ at $z=10$ and increasing with decreasing redshift.
As just mentioned, this trend is due to the adopted fit in eq.~\ref{eq:fit}, which favors more massive PopII-dominated galaxies.
On the contrary, temperature selected PopII-dominated DLAs feature fractions of just 20\% at $z=10$ and 60\% at $z=5.5$, as more massive galaxies are excluded by the temperature selection criterion.

From these results, we also conclude that the fraction of intermediate DLAs is rather low for both definitions.
Precisely, it is $<4\%$ in the VMSN and MSN models and only slightly larger in the RSN model (see also Fig.~\ref{tvsmgas} and Fig.~\ref{lyaxvx}).

While the fraction of PopIII-dominated DLAs becomes larger at higher redshifts, the total DLA number density is lower at earlier epochs, because less structures are present.
Thus, to predict how many high-$z$ DLAs are observable as PopIII-dominated, we calculate the expected cumulative fraction of all DLAs at $z \ge 5.5$ that are PopIII-dominated.
For temperature selected DLAs, it is computed by
\begin{equation}
f_{{\rm DLA},T} = \frac{\int_{z>5.5} \mathrm{d}z' ~N_{i}(z') \Phi(z') \frac{\mathrm{d}V}{\mathrm{d}z'}} 
                       {\int_{z>5.5} \mathrm{d}z' ~N_{\rm tot}(z')  \Phi(z') \frac{\mathrm{d}V}{\mathrm{d}z'}},
\end{equation}
where $\Phi(z') \propto 10^{-0.5*z'}$ denotes 
the quasar number at redshift larger than $z'$ \citep{Fan2001},
$\mathrm{d}V/\mathrm{d}z' = D^{2} \frac{\mathrm{d}D}{\mathrm{d}z'} \mathrm{d}\Omega$ indicates the comoving cosmic volume and $D(z')$ is the comoving distance between redshift $z'$ and $0$.
The cumulative fraction is weighted by the number of background quasars, because  the number of observable DLAs also depends on the number of background sources \citep{Fan2001}.
As a result, we get values of $42.5\%$, $43.6\%$ and $36.5\%$ for the VMSN, MSN and RSN model, respectively.
For mass selected DLAs, this cumulative fraction is expressed as
\begin{equation}
f_{{\rm DLA},M} = \frac{\int_{z>5.5}  \mathrm{d}z' ~\Delta_{i}(z') \Phi(z') (1+z')^{2} \frac{\mathrm{d}D}{\mathrm{d}z'}} 
                       {\int_{z>5.5}  \mathrm{d}z' ~\Delta_{\rm tot}(z') \Phi(z') (1+z')^{2} \frac{\mathrm{d}D}{\mathrm{d}z'}}, 
\end{equation}
which gives $3.4\%$, $3.5\%$ and $2.6\%$ for VMSN, MSN and RSN, respectively.
%
%
\subsection{DLA properties}
%
\begin{figure*}
\centering
\includegraphics[width=0.90\linewidth]{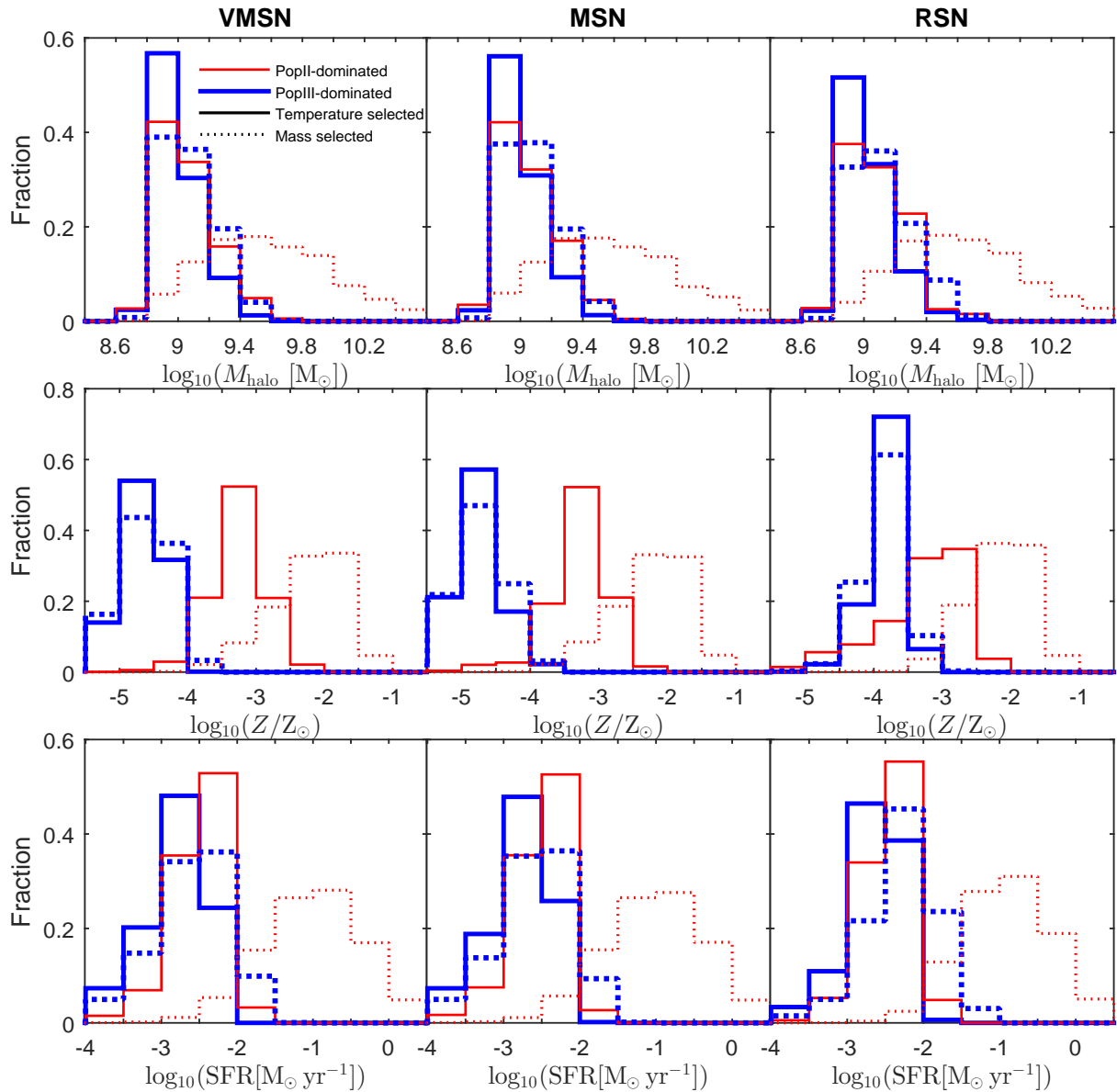}
\caption{
From top to bottom, halo mass [$\rm M_\odot$], metallicity [$\rm Z_\odot$] and SFR [$\rm M_\odot~yr^{-1}$] distributions of PopII-dominated (red thin lines) and PopIII-dominated (blue thick lines) DLAs at $z\ge5.5$, in base-10 logarithmic scale.
Solid lines indicate temperature selected DLAs, while dotted lines indicate mass selected DLAs.
From left to right, the columns refer to model VMSN, MSN and RSN.
\newline
(The color version is only available in the online journal.)
}
\label{fracvsme}
\end{figure*}
In Fig.~\ref{fracvsme} we show the distributions of halo mass, metallicity and SFR of PopIII-dominated (blue thick lines) and PopII-dominated (red thin lines) DLAs at $z\ge5.5$.
For temperature selected DLAs (solid lines) the fraction of class $i$ DLAs in the $j$-th bin is computed by
\begin{equation}
F_{i,j;T} = \frac{\int_{z>5.5} \mathrm{d}z' N_{i,j}(z') \Phi(z') \frac{\mathrm{d}V}{\mathrm{d}z'}}
                 {\int_{z>5.5} \mathrm{d}z' N_{i, \rm tot}(z') \Phi(z') \frac{\mathrm{d}V}{\mathrm{d}z'}},
\end{equation}
where $N_{i,j}(z')$ is the number of class $i$ DLAs in the $j$-th bin at redshift $z'$, and $N_{i, \rm tot}(z')$ is the total number of class $i$ DLAs at $z'$.
For mass selected DLAs (dotted lines), it is calculated by
\begin{equation}
F_{i,j;M} = \frac{\int_{z>5.5} \mathrm{d}z' \Delta_{i,j}(z') \Phi(z') (1+z')^{2} \frac{\mathrm{d}D}{\mathrm{d}z'} }
                 {\int_{z>5.5} \mathrm{d}z' \Delta_{i, \rm tot}(z') \Phi(z')  (1+z')^{2} \frac{\mathrm{d}D}{\mathrm{d}z'}},
\end{equation}
where $\Delta_{i,j}(z')$ is the sum of $\Sigma(M_{\rm halo})$ of class $i$ DLAs in the $j$-th bin at $z'$, and $\Delta_{i, \rm tot}(z')$ is the sum of $\Sigma(M_{\rm halo})$ of all class $i$ DLAs at $z'$.

Although no obvious dependence on the first star models, the properties of PopII-dominated DLAs (red lines) feature major differences according to the selection criteria.
Specifically, temperature selected PopII-dominated DLAs (solid red lines) have very low halo masses $M_{\rm halo}<10^{9.6}\,\rm M_{\odot}$, while metallicities are concentrated in the range $(10^{-4.5}-10^{-2})\,\mathrm{Z}_{\odot}$ and peak around $Z\approx10^{-3}\,\mathrm{Z}_{\odot}$.
Their SFRs are $\rm <10^{-1.5}~M_{\odot}\,yr^{-1}$.
Since eq.~\ref{eq:fit} suppresses low-mass structures and imposes a roughly flat trend of neutral hydrogen cross-section for big structures, the mass selected PopII-dominated DLAs (dotted red lines) have flatter halo mass distributions with $M_{\rm halo}>10^{8.8}\,{\rm M}_{\odot}$ and a peak at $\approx 10^{9.6}\,{\rm M}_{\odot}$.
Their metallicities are in the range ($10^{-3.5} - 10^{-1})\,\mathrm{Z}_{\odot}$ and peak at $\approx 10^{-2}\,\mathrm{Z}_{\odot}$, which is 1~dex higher than those of temperature selected ones.
They also show larger SFRs ($> 10^{-2.5}\,\rm M_{\odot}\,yr^{-1}$), that peak at $\approx 0.1\,\rm M_{\odot}\,yr^{-1}$.

As the two DLA definitions select similar samples of PopIII-dominated galaxies (see Fig.~\ref{tvsmgas}), the distributions of PopIII-dominated DLAs (blue lines) show similar trends, although mass selected ones have slightly higher mass, metallicity and SFR.
They are typically hosted in halos with mass
$M_{\rm halo} < 10^{9.6}~\rm M_{\odot}$ (also see Fig.~\ref{tvsmgas}), consistently with \cite{Kulkarni2013ApJ} and \cite{MT2015}, and peak at $\approx 10^9~\rm M_\odot$.
Their typical metallicities are, as expected, lower than those of PopII-dominated DLAs, i.e. $Z<10^{-4}~\rm Z_\odot$ in the VMSN and MSN models, while $10^{-5}~\rm Z_\odot < Z < 10^{-3}~\rm Z_\odot$ in the RSN model, 
since the higher small-mass PopIII SFR in the RSN model induces more metal pollution in low-mass galaxies.
Their SFRs are $< 10^{-1.5}\,\rm M_{\odot}\, yr^{-1}$ and peaks around $\approx 10^{-2.5}\,\rm M_{\odot}\, yr^{-1}$. 

Temperature selected PopII-dominated (solid red lines) DLAs present halo mass and SFR distributions very similar to those of the PopIII-dominated ones (solid blue lines), while they still can be neatly identified through metallicities.
In fact, most PopIII-dominated DLAs have
$Z <10^{-3.5}\,\mathrm{Z}_{\odot}$, while only a few temperature selected PopII-dominated DLAs reach such low values.

\section{DLA Metal Abundances}
\label{sec:metal}

In this section, we present the results for metal abundance ratios of indicative heavy elements extracted from our simulations.
We focus on carbon (C), oxygen (O), silicon (Si) and iron (Fe), which are abundant in galaxies and have been detected in high-$z$ DLAs \citep[][]{Becker2012, Morrison2016}.

The abundance ratios with respect to the corresponding solar values are defined as
[A/B] = log$_{10}$($N_{\rm A}$/$N_{\rm B}$) $-$ log$_{10}(N_{\rm A}/N_{\rm B})_{\odot}$,
where A and B are two arbitrary species, $N_{\rm A(B)}$ is the number densities of element A(B) and the subscript $\odot$ denotes the solar values from \cite{Asplund2009}.

%
\subsection{Abundance ratios}
\label{sec:metal:galaxies}
\begin{figure*}
\centering
\includegraphics[width=0.9\linewidth]{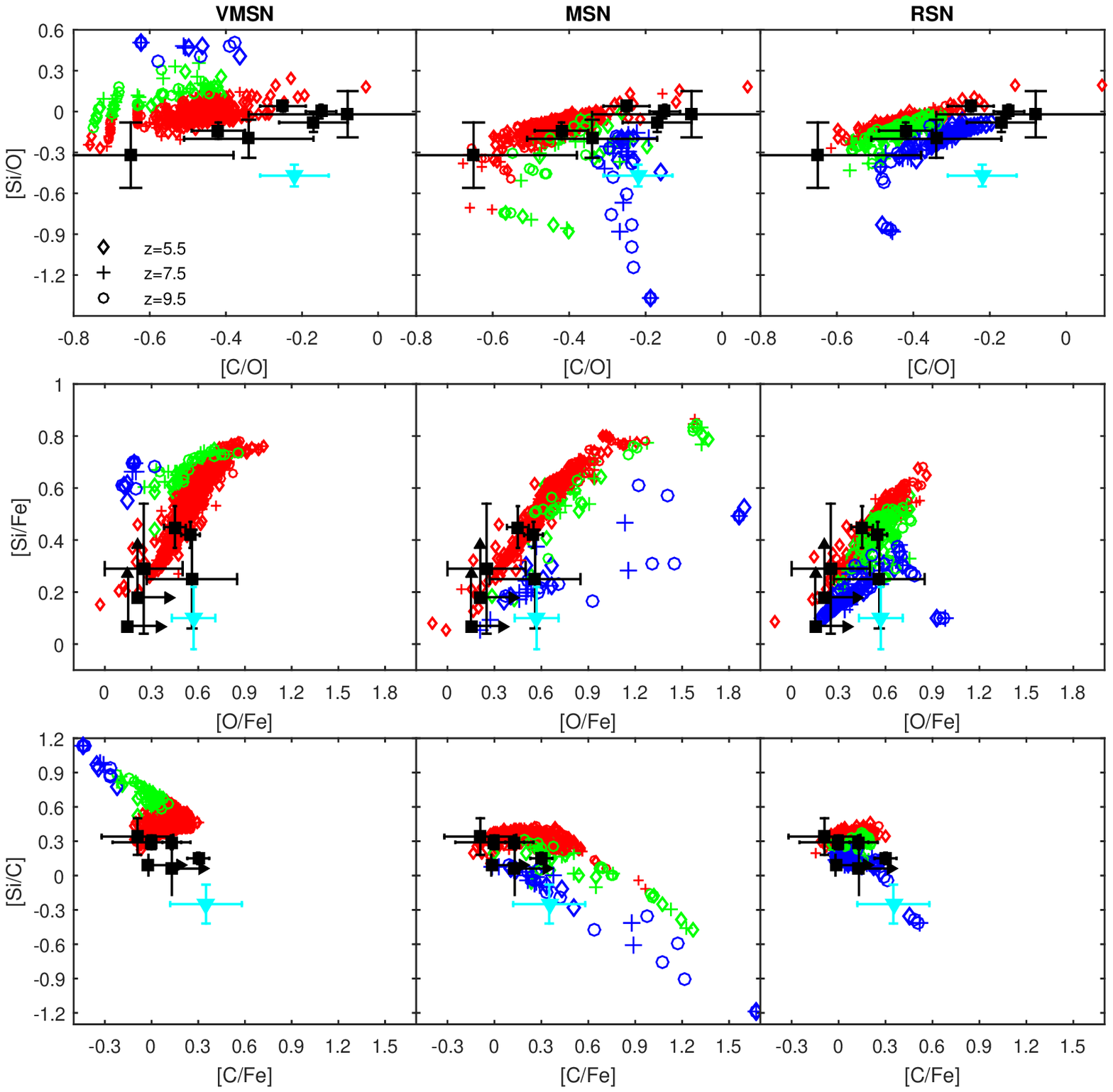}
\caption{
Distributions of abundance ratios [Si/O] vs [C/O] (top panels), [Si/Fe] vs [O/Fe] (central) and [Si/C] vs [C/Fe] (bottom).
From left to right the columns refer to models: VMSN, MSN and RSN. The symbols denote PopII-dominated (small red), intermediate (green) and PopIII-dominated (big blue) galaxies at $z=9.5$ (circles), 7.5 (crosses) and 5.5 (diamonds).
Black solid squares with error bars (or arrows to denote lower or upper limits) are DLAs observed in the redshift range $5.3 - 6.2$ (\citealt{Becker2012}), while the cyan solid triangle refers to a sub-DLA at $z=5$ reported by \citet{Morrison2016}.
\newline
(The color version is only available in the online journal.)
}
\label{lyaxvx}
\end{figure*}
In Fig.~\ref{lyaxvx} we show distributions of abundance ratios at redshift $z = 9.5$, 7.5 and 5.5 for, respectively, 617, 1496 and 2581 metal enriched galaxies in the VMSN model; 608, 1508 and 2590 in the MSN model; and 512, 1332 and 2370 in the RSN model. 
The RSN model features fewer (by $\sim 16\%$ at $z = 9.5$, $\sim 11\%$ at $z = 7.5$ and $\sim 8\%$ at $z = 5.5$) metal polluted galaxies than the other models due to the longer lifetimes of small-mass first stars that delay the initial phases of cosmic metal enrichment.
All these galaxies are possible DLA candidates when selected according to the mass criterion,
while only 10\%-30\% of them can be DLAs when selected according to their gas temperature.
We have verified that this smaller sample of DLAs retains the same characteristics of the metal ratio distributions of the full sample. This is because the metal abundance ratios of galaxies are mainly determined by the yields of the polluting stellar population and thus are weakly dependent on the halo mass or gas temperature.
For this reason, the distributions are representative of the DLA population, independently from the selection.
In the remaining of the paper we will thus refer to abundance ratios of DLAs rather than of galaxies.

From Fig.~\ref{lyaxvx}, PopII-dominated, intermediate and PopIII-dominated DLAs are clearly separated in the panels [Si/O] vs [C/O], [Si/Fe] vs [O/Fe] and [Si/C] vs [C/Fe] for the VMSN and MSN models, while in the RSN model they mostly overlap.

More specifically, PopII-dominated DLAs (red symbols) display abundance ratios very similar in the three models, indicating their common origin (i.e. pollution by the same PopII generation).
A linear correlation between [Si/O] and [C/O] is evident, due to the different metal pollution phase in each DLA. 
The [Si/O] ratios of PopII-dominated DLAs are mostly between $-0.2$ and $0.2$, while typically [C/O] ratios are in the range $[-0.6,-0.3]$.
A few cases at $z=5.5$ have [C/O] $>-0.3$ due to pollution by local AGB stars, whose contribution becomes relevant at such epochs as a consequence of AGB lifetimes\footnote{
AGB stars evolve in a few $ 10^8\,\rm yr$, hence they can enrich the medium already at $z\sim 6$ by spreading metals via mass loss.
}.
PopII-dominated DLAs also show a roughly linear relation between [Si/Fe] and [O/Fe] and they are preferentially located in regions where [O/Fe]~$=[0.3,0.9]$ and [Si/Fe]~$= [0.2,0.7]$.
Some have very low [O/Fe] and [Si/Fe] ratios at $z=5.5$, resulting from metal spreading from stars dying as SNIa (with high iron yields) that become increasingly important with decreasing redshift.
In the [Si/C] vs [C/Fe] panels, PopII-dominated DLAs in the three models share the region delimited by [C/Fe]~$= [-0.1, 0.3]$ and [Si/C]~$= [0.2, 0.5]$.
Some differences appear, though, e.g. DLAs with [Si/C]~$> 0.5$ are present only in the VMSN model as a result of the high [Si/C] yields ($\approx 1.14$) from PISN.
Then, only the MSN model features some PopII-dominated DLAs with [C/Fe]~$>0.3$ because of the low iron yields expected for faint supernovae from massive stars \citep{Heger2010}.

The distributions of PopIII-dominated DLAs (blue symbols) are visibly different in the three models, and do not show obvious time evolution, because of the
leading role of the metal yields for massive stars at these epochs.
We note that a PopIII IMF \citep{Hirano2015} changing with redshift
and having strongly variable yields for massive stars might lead to an evolution of the distributions for PopIII-dominated DLAs.

Thanks to the large explosion energy of PISN and the short lifetime of very massive PopIII stars, PopIII-dominated DLAs in the VMSN model are mostly concentrated in a small region of the [Si/O] vs [C/O] panel with [C/O]~$\approx - 0.62$ and [Si/O]~$\approx 0.51$.
Similarly, these DLAs typically have [O/Fe]~$\approx 0.18$ vs [Si/Fe]~$\approx 0.7$ and [C/Fe]~$\approx -0.44$ vs [Si/C]~$\approx 1.14$.
Although their [C/O] and [O/Fe] might be similar to those of PopII-dominated and intermediate DLAs, they would display higher [Si/O] and [Si/Fe], e.g. [Si/O]~$> 0.3$.

Since the MSN model includes PopIII metal enrichment both from massive stars ($\rm > 40\,M_{\odot}$), that have very high carbon and oxygen but low silicon and iron yields when exploding as faint supernovae, and from regular stars ($\rm <40\,M_{\odot}$), whose metal yields are very similar to those of PopII stars, a number of PopIII-dominated DLAs have [Si/O]~$\approx -1.37$ and 
[C/O]~$\approx -0.19$,
and some are closer to PopII-dominated DLAs, while others are located in between, as a result of the combined contribution of massive and low-mass PopIII stars.
The same effects also appear in the distribution of [Si/Fe] vs [O/Fe], i.e. some of them have [O/Fe] $\approx 1.9$ and [Si/Fe] $\approx 0.5$, while others are located around [O/Fe] $\approx 0.5$ and [Si/Fe] $\approx 0.2$.
In the [Si/C] vs [C/Fe] panel, they are distributed in the range [C/Fe]~$= [0,1.68]$ and [Si/C]~$= [-1.2,0]$.

The metal yields of pristine SNe with small-mass progenitors are similar to those from PopII/I stars, thus PopIII-dominated DLAs in the RSN model are located near PopII-dominated and intermediate DLAs, although some differences are still visible.
For example, given the same [C/O] (or [O/Fe]), they present [Si/O] (or [Si/Fe]) $\approx 0.2$ dex lower than that of PopII-dominated DLAs, and they also feature a smaller [Si/C] ($< 0.2$).

Intermediate DLAs (green symbols) always locate between PopII-dominated and PopIII-dominated ones.
They are relatively few in the VMSN and MSN models when compared to the RSN model, as mentioned earlier.
Since massive PopII stars (with mass $>30\,\rm M_{\odot}$) have stellar lifetimes much shorter than low-mass ones, intermediate DLAs retain some hints of their metal pollution by having [C/O] ratios lower, but [Si/Fe] ratios higher than the PopIII-dominated DLAs, which are the typical features of PopII stars with mass $>30\,\rm M_{\odot}$.

The unique characteristics shown by PopIII-dominated DLAs in the abundance ratios can be exploited to identify first star signatures and constrain PopIII mass ranges by comparing them to observed DLAs at $z\ge5$ with high precision abundance ratios measured (\citealt{Becker2012, Morrison2016}).
These are listed in Table~\ref{data_dla} and shown as symbols in Fig.~\ref{lyaxvx}.
\begin{table*}
\centering
\begin{tabular}{c c c c c c c c}
\hline \hline
QSO             &  $z$      & [Si/O]           & [C/O]            & [Si/C]           & [C/Fe]           & [O/Fe]          & [Si/Fe] \\
\hline
SDSS J1202+3235 & 5.0    & -0.47 $\pm$ 0.08 & -0.22 $\pm$ 0.09 & -0.25 $\pm$ 0.17 & 0.35 $\pm$ 0.23  & 0.57 $\pm$ 0.14 & 0.10 $\pm$ 0.12 \\
SDSS J0231-0728 & 5.3380 & -0.14 $\pm$ 0.06 & -0.42 $\pm$ 0.07 & 0.29 $\pm$ 0.06  & 0.13 $\pm$ 0.06  & 0.55 $\pm$ 0.06 & 0.42 $\pm$ 0.05 \\
SDSS J0818+1722 & 5.7911 & 0.00 $\pm$ 0.05  & -0.15 $\pm$ 0.04 & 0.15 $\pm$ 0.05  & 0.30 $\pm$ 0.07  & 0.45 $\pm$ 0.07 & 0.45 $\pm$ 0.08 \\
SDSS J0818+1722 & 5.8765 & -0.08 $\pm$ 0.07 & -0.17 $\pm$ 0.09 & 0.09 $\pm$ 0.09  & > - 0.02         & >0.15           & >0.07           \\
SDSS J1148+5251 & 6.0115 & 0.04 $\pm$ 0.04  & -0.25 $\pm$ 0.06 & 0.29 $\pm$ 0.06  & -0.00 $\pm$ 0.25 & 0.25 $\pm$ 0.25 & 0.29 $\pm$ 0.25 \\
SDSS J1148+5251 & 6.1312 & -0.32 $\pm$ 0.24 & -0.65 $\pm$ 0.27 & 0.34 $\pm$ 0.16  & -0.09 $\pm$ 0.23 & 0.56 $\pm$ 0.29 & 0.25 $\pm$ 0.19 \\
SDSS J1148+5251 & 6.1988 & -0.20 $\pm$ 0.14 & -0.34 $\pm$ 0.17 & 0.14 $\pm$ 0.12  &  ...             &  ...            &  ...            \\
SDSS J1148+5251 & 6.2575 & -0.02 $\pm$ 0.17 & -0.08 $\pm$ 0.26 & 0.06 $\pm$ 0.23  & >0.13      & >0.21  & >0.18  \\
\hline
\end{tabular}
\caption{
Abundance ratios of DLAs observed at $z\ge5$. The first measurement is from \citet{Morrison2016}, while the others are from \citet{Becker2012}.
}
\label{data_dla}
\end{table*}

Most of data points are consistent with PopII-dominated DLAs, although some of them have error bars too large to draw definitive conclusions.
Meanwhile, some data points seem to favor the MSN and RSN models, although no observation can clearly discriminate one model over the others.

The DLA at $z=5.8765$ found along the line-of-sight of QSO SDSS~J0818+1722 is very close to the PopIII-dominated DLAs in the RSN model, as indicated by the [Si/O] vs [C/O] panel, as well as by the ratio
[Si/C]$=0.09 \pm 0.09$.
Nevertheless, there are only lower limits for its [Si/Fe] and [O/Fe], which can not then be used to set more stringent constraints on its class and progenitors.

The DLA at $z=5.7911$ is also consistent with PopIII-dominated objects in the RSN model in the [Si/O] vs [C/O] panel, while it clearly locates in the region of PopII-dominated DLAs in the [Si/Fe] vs [O/Fe] panel.

All the abundance ratios of the recently observed sub-DLA at $z=5$ from QSO SDSS~J1202+3235 \cite[][]{Morrison2016} are consistent with PopIII-dominated DLAs in the MSN model, although it does not display the features of faint supernovae, e.g. very low [Si/O] ($<-1$) or very high [C/Fe] ($>1$).
Its abundance ratios are also close to those of PopIII-dominated DLAs in the RSN model in the panel [Si/C] vs [C/Fe], but they are not fully consistent with the same model in the panel [Si/O] vs [C/O].
We also note that the metallicity of this sub-DLA ([O/H] $=-2.00 \pm 0.12$) is much higher than the PopIII-dominated DLAs predicted in our simulations. 

%
\subsection{Evolution of DLA mean abundance ratios}
\label{sec:metal:mean}
%
\begin{figure}
\centering
\includegraphics[width=0.90\linewidth]{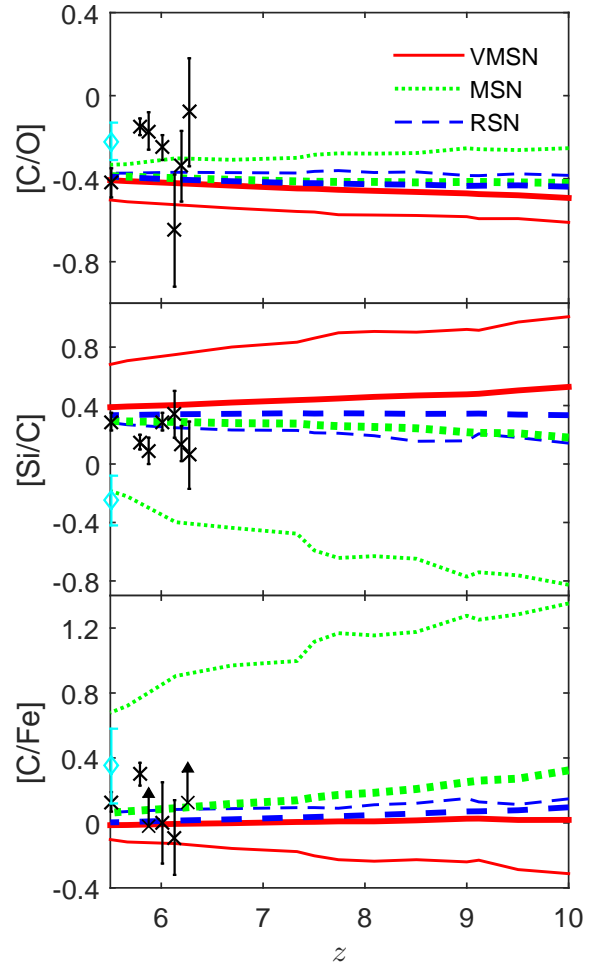}
\caption{
Mean value of [C/O] (top), [Si/C] (middle) and [C/Fe] (bottom) in the VMSN (solid red lines), MSN (dotted green) and RSN (dashed blue) models as a function of redshift. The thin and thick curve sets denote temperature selected and mass selected DLAs, respectively. The black crosses with error bars (or arrows) are data for observed DLAs taken from \citet{Becker2012}, and the one showed as cyan diamond with error bar is from \citet{Morrison2016}.
\newline
(The color version is only available in the online journal.)
}
\label{xxvsz}
\end{figure}
In Fig.~\ref{xxvsz}, we show the evolution of the average [C/O], [Si/C] and [C/Fe] for DLAs selected under both definitions.
As mass selected DLAs are dominated by massive halos with metal abundances predominantly coming from PopII stars, their mean abundance ratios (thick lines) are weakly dependent on the first star models and present a mild evolution with redshift \citep{Kulkarni2013ApJ}.
Instead, the mean abundance ratios of temperature selected DLAs (thin lines) are dramatically different in the three models and show an evident redshift evolution.

Specifically, since PopIII stars in all three models have [C/O] yields similar to those of PopII stars, the mean [C/O] ratios of DLAs present differences of only 0.3~dex for the temperature selected ones and 0.1~dex for the mass selected ones.
In most cases, the mean [C/O] have a little increase with decreasing redshift, reaching $\approx -0.4$ at $z=5.5$, as a result of the increasing contribution from AGB stars.
In the VMSN model, this behaviour is also partly caused by the low [C/O] yields of PISN ($<-0.6$).
PISN pollution is mostly significant at high redshift, so at $z=10$ the mean [C/O] in VMSN model is $\approx-0.6$ ($\approx-0.5$) for temperature (mass) selected DLAs.
However, temperature selected DLAs in the MSN model show an exception, i.e. their average [C/O] decreases with decreasing redshift. 
Indeed, massive first stars, that explode as faint supernovae with high [C/O] yields ($\approx-0.2$), determine a relatively high mean [C/O] ratio at $z=10$. As redshift decreases, the mean [C/O] is gradually reduced by the growing contribution of PopII stars that have lower [C/O] yields ($\approx-0.4$).

PopIII-dominated DLAs have very high (low) [Si/C] in the VMSN (MSN) (see also bottom panels in Fig.~\ref{lyaxvx}), thus the average [Si/C] values feature the largest differences within the three models, e.g. $\approx1.8$ dex ($\approx0.4$ dex) for temperature selected (mass selected) DLAs at $z=10$.
As redshift decreases, the contribution from PopIII stars becomes less important, hence the mean [Si/C] ratios in both VMSN and MSN models tend to converge to that of the RSN model, with [Si/C]~$\approx 0.3$.

The mean [C/Fe] ratio in the RSN model is $\approx 0$, and slowly decreases with decreasing redshift. 
This might be caused by SNIa events that start contributing to metal enrichment at low redshift and injecting important amounts of iron.
Although mass selected DLAs in the VMSN model have mean [C/Fe] very similar to that of the RSN model, as a result of their small fraction of PopIII-dominated DLAs (see Fig.~\ref{lyafvz}), the temperature selected ones in this model show a low mean [C/Fe] at $z=10$ ($\approx-0.3$) because of the low [C/Fe] yields of PISN,  while it increases with decreasing redshift and finally converges to the RSN trend.
With larger [C/Fe] yields from faint SNe, the average [C/Fe] in the MSN model is the highest, e.g. [C/Fe]~$\approx 1.3$ ($\approx 0.3$) for temperature selected (mass selected) DLAs at $z=10$.

Despite [C/O] measurements would not help discriminate the various PopIII models in a solid way, it is possible to infer interesting conclusions from observational data of [Si/C] and [C/Fe].
The predicted mean metal ratios of mass selected DLAs (thick lines), for all three models, are consistent with the observed high-$z$ DLAs \cite[see also ][]{Kulkarni2013ApJ}, while they have clear difficulties in reproducing the observation by \citet{Morrison2016} for [Si/C] and [C/Fe].
Temperature selected DLAs (thin lines) feature trends similar to the ones obtained by mass selection in the RSN model, while they show higher (lower) mean [Si/C] in the VMSN (MSN) model, as well as much higher [C/Fe] in the MSN model.
These latter trends are in tension with [Si/C] and [C/Fe] data by \citet{Becker2012}, but explain reasonably well the data of \citet{Morrison2016} through MSN enrichment.
We note that the various differences may be caused by a bias in the observations, e.g. PopIII-dominated DLAs with such low metallicities ($Z<10^{-3.5}\,\rm Z_{\odot}$) might not be easy to detect \citep{Wolfe2005}.
%
%
\section{Discussion and Conclusions}
\label{sec:discussion}
%
%
We use hydrodynamical chemistry simulations with different PopIII IMFs to study the metal signatures of the first stars in high-$z$ ($z \ge 5.5$) DLAs.
The simulations include atomic and molecular chemistry, stellar evolution and metal enrichment according to metal-dependent yields and lifetimes.
In all simulations, PopII/I stars are assumed to be distributed according to a Salpeter IMF with mass range [0.1, 100]~$\rm M_{\odot}$ and include metal pollution from type II SNe, AGBs and type Ia SNe.
Three cases for PopIII stars have been considered:
(i) a top-heavy IMF with mass range [100, 500]~$\rm M_{\odot}$ and PISN explosions of stars with mass [140, 260]~$\rm M_{\odot}$ (VMSN);
(ii) a normal Salpeter IMF, with mass range [0.1, 100]~$\rm M_{\odot}$, which includes massive SNII explosions within [10, 100] $\rm M_{\odot}$ (MSN);
and 
(iii) a normal Salpeter IMF with SNII explosions within [10, 40]~$\rm M_{\odot}$ (RSN).
Thanks to our detailed implementation, we can retrieve solid results for the chemistry and the pollution history of the primordial Universe, which are mainly led by the first PopIII and PopII/I star formation episodes and by the interplay between different feedback mechanisms.

Although our simulations cover a wide range of PopIII IMFs, other options could be possible for PopIII stars, while even the most extreme scenarios are unlikely to have first stars above 500~$\rm M_\odot$.
Similarly, stellar structure models and physical processes in the stellar cores different from those adopted here, e.g. explosion mechanisms, magnetic fields, rotation and nuclear reaction rates, might affect the theoretical metal yields from supernova explosions and thus the predicted metal abundance ratios for DLAs \citep[see more discussion in][]{Ma2017}, nevertheless the cosmic gas evolution would not be changed significantly.
The PopIII SFR, and thus the corresponding metal contribution, may be slightly changed by the adopted criteria for the transition from a PopIII to a PopII/I star formation mode, but the PopII/I SFR is hardly affected \cite[][]{Maio2010}.
While significant changes are not expected when varying the adopted underlying cosmological model, effects from warm dark matter \cite[][]{MV2015} can suppress the early SFR, and primordial streaming motions \cite[][]{TH2010, Maio2011vb} can moderately impact gas evolution and primordial gas collapse at very high $z$.

In our analysis, we explore the patterns of DLAs extracted from the numerical simulations with two possible selection criteria: the first one selects DLAs as gaseous objects with mass-weighted temperature $ T< 10^{4}\,\rm K$ (temperature selected DLAs; \citealt{Cooke2015}), and the other one is based on the associated neutral hydrogen cross-section (mass selected DLAs; as in \citealt{Kulkarni2013ApJ}).
Our main results can be summarized as follows.
\begin{itemize}
\item The fraction of DLAs that are PopIII-dominated decreases quickly with redshift, almost independently from the first star models.
However, it is dramatically affected by the selection criteria for DLAs, as it can be as high as $40\%$ at $z=6$ for temperature selected DLAs, but only $3\%$ for mass selected ones.
\item PopIII-dominated DLAs have very low halo mass ($<10^{9.6}\rm M_{\odot}$),
metallicities ($<10^{-4}\rm Z_{\odot}$ in the VMSN and MSN model and $<10^{-3}\rm Z_{\odot}$ in the RSN model) 
and SFR ($<0.01 \rm M_{\odot}\, yr^{-1}$),
without much dependence on the selection criteria.
\item PopII-dominated DLAs, instead, display different properties under the two adopted definitions.
While temperature selected PopII-dominated DLAs show halo mass and SFR distributions very similar to those of PopIII-dominated ones and $\approx 1$ dex larger metallicities, mass selected PopII-dominated DLAs span a larger mass range, i.e. $(\rm 10^{8.6} - 10^{10.6})~M_{\odot}$, and favor larger metallicities ($>10^{-3.5} \,\rm Z_{\odot}$) and SFRs ($>10^{-2.5} \,\rm M_{\odot}\,yr^{-1}$).
\item No matter which definition we adopt, the PopIII-dominated DLAs could be easily identified through their abundance ratios in the VMSN and MSN models (and possibly also in the RSN model), e.g. by comparing [Si/O] vs [C/O], [Si/Fe] vs [O/Fe] or [Si/C] vs [C/Fe].
Besides, the average abundance ratios of DLAs (e.g. [C/O], [Si/C] and [C/Fe]) from all the simulations display a visible difference at high redshift, but converge quickly with decreasing redshift.
\item While most currently observed DLAs at $z\ge 5$ \citep{Becker2012, Morrison2016} have metal ratios consistent with a PopII stellar enrichment, the one presented in \cite{Morrison2016} can be explained only with the MSN model and at least one DLA presented in \cite{Becker2012} can be well described by the RSN model. None though supports the VMSN model.
Despite more precise observations are necessary in the future, these results seem consistent with the conclusions drawn from the most metal-poor DLA currently known \cite[][]{Cooke2017}, which
{ feature no significant signatures from massive first stars} 
and point towards enrichment from lower-mass stars to explain observed low-metallicity gas patterns.
\end{itemize}

\section*{Acknowledgments}
We thank the anonymous referee for careful reading and useful discussions that improved the presentation of the results.
We used the tools offered by the NASA Astrophysics Data Systems  and by the JSTOR archive for bibliographic research.
Q. Ma is partially supported by the National Natural Science Foundation of China (Grant No. 11373068 and No. 11322328), the National Basic Research Program ("973" Program) of China (Grants No. 2014CB845800 and No. 2013CB834900), and the Strategic Priority Research Program "The Emergence of Cosmological Structures" (Grant No. XDB09000000) of the Chinese Academy of Sciences.

\bibliography{ref}
\appendix

\end{document}